<u>Original Paper</u>

# Searching COVID-19 Clinical Research Using Graph Queries: Algorithm Development and Validation

Francesco Invernici, MSc; Anna Bernasconi, PhD; Stefano Ceri, PhD

Department of Electronics, Information, and Bioengineering, Politecnico di Milano, Milan, Italy

**Corresponding Author:**
Anna Bernasconi, PhD
Department of Electronics, Information, and Bioengineering
Politecnico di Milano
Via Ponzio 34/5
Milan, 20133
Italy
Phone: 39 23993494
Fax: 39 23993411
Email: anna.bernasconi@polimi.it

## *Abstract*

**Background:** Since the beginning of the COVID-19 pandemic, >1 million studies have been collected within the COVID-19 Open Research Dataset, a corpus of manuscripts created to accelerate research against the disease. Their related abstracts hold a wealth of information that remains largely unexplored and difficult to search due to its unstructured nature. Keyword-based search is the standard approach, which allows users to retrieve the documents of a corpus that contain (all or some of) the words in a target list. This type of search, however, does not provide visual support to the task and is not suited to expressing complex queries or compensating for missing specifications.

**Objective:** This study aims to consider small graphs of concepts and exploit them for expressing graph searches over existing COVID-19–related literature, leveraging the increasing use of graphs to represent and query scientific knowledge and providing a user-friendly search and exploration experience.

**Methods:** We considered the COVID-19 Open Research Dataset corpus and summarized its content by annotating the publications' abstracts using terms selected from the Unified Medical Language System and the Ontology of Coronavirus Infectious Disease. Then, we built a co-occurrence network that includes all relevant concepts mentioned in the corpus, establishing connections when their mutual information is relevant. A sophisticated graph query engine was built to allow the identification of the best matches of graph queries on the network. It also supports partial matches and suggests potential query completions using shortest paths.

**Results:** We built a large co-occurrence network, consisting of 128,249 entities and 47,198,965 relationships; the GRAPH-SEARCH interface allows users to explore the network by formulating or adapting graph queries; it produces a bibliography of publications, which are globally ranked; and each publication is further associated with the specific parts of the query that it explains, thereby allowing the user to understand each aspect of the matching.

**Conclusions:** Our approach supports the process of query formulation and evidence search upon a large text corpus; it can be reapplied to any scientific domain where documents corpora and curated ontologies are made available.

## *Introduction*

Since the COVID-19 pandemic outbreak in early 2020, important clinical research efforts have been targeted at understanding the COVID-19 disease. More than 1 million studies have been collected within the COVID-19 Open Research Dataset (CORD-19), a corpus of manuscripts created to accelerate the research against the disease. Their related abstracts hold a wealth of information that remains largely unexplored and difficult to search due to its unstructured nature.

Searching over the literature is a nontrivial task, as it strongly relies on the quality of the data corpus, the characteristics of the search portal, and the language used to express the search.

Keyword-based search is the standard search approach, which allows users to retrieve the documents of a corpus that contain some of the words in a specified target list [1,2]. However, this type of search lacks visual support for the task and is not suitable for expressing complex research queries or compensating for missing specifications.

The development of frontend tools and visualizations for COVID-19 knowledge graphs has been motivated by several works [3,4]. We then explored the use of small graph-based queries that can be built visually [5] to empower a literature exploration tool; the GRAPH-SEARCH system stems from this motivation, providing both a visual language to express search queries and a friendly tool to explore relevant publications, which highlights the relationships between the original graph queries and an underlying corpus of scientific evidence, in the spirit of literature-based discovery [6].

To support this idea, the underlying textual corpus must first be analyzed and enriched; in our approach, the CORD-19 was expressed in the form of a co-occurrence network. First, we annotated all the abstracts with terms from the Unified Medical Language System (UMLS) [7] and the Coronavirus Infectious Disease Ontology (CIDO) [8]. This step closely aligns with classical work on ontology-based annotation (refer to Semantic MEDLINE [9] and our previous study on genomic metadata annotation [10,11]). Second, we built a comprehensive co-occurrence network that includes all relevant clinical and biological concepts mentioned in the corpus, linking them based on their co-occurrence in given abstracts.

The visual language used to express a query over the network describes concepts as nodes and their copresence within research abstracts as undirected edges; some concepts are associated with medical conditions, whereas others are associated with treatments or biological entities. We also allow modifiers. Queries run on the network may correspond to the expressed graph pattern or to a selected subpart.

The query semantics corresponds to extracting scientific evidence (ie, publications) from the corpus, in support of the existence of the relationships linking the expressed concepts; each search process extracts the references that best explain the relationships occurring within the query. When a specified path is not present in the co-occurrence network, alternative scored and ranked shortest paths connecting the nodes expressed in the query are proposed to the user (refer to the *Methods* section). The search output provides a ranking of references because of their weight, summing up the support that they provide to several relationships in the query.

Our GRAPH-SEARCH implementation is supported by a graphical interface (refer to the *Data Availability* section) that allows the user to express the queries and to interpret the results in terms of concepts explained by each discovered reference, thus enabling the users to better qualify the query during the interaction; in addition, users can read the textual abstracts of the retrieved references. Such interactive exploration of the search space allows for exploring assumptions and for progressively adapting them as a result of existing evidence.

The manuscript is organized as follows: we first describe the CORD-19, the characteristics of the co-occurrence network representing CORD-19 abstracts, the technological process of building the network, the graph search operation, and the web user interface that allows us to express graph queries and explore the retrieved results. We then present a series of example use case (UC) queries relevant to COVID-19 research and review the current state of the art. We then evaluate the benefits of using our GRAPH-SEARCH as opposed to full-text indexed databases and keyword search. Finally, we draw our conclusions.

## Methods

### The CORD-19 Corpus

CORD-19 [12] is a corpus of academic publications about COVID-19 and related coronavirus research; it was released and maintained by the Allen Institute for AI in collaboration with The White House Office of Science and Technology Policy and other partners. Published articles and preprints were collected from several archives, including PubMed, PubMedCentral, bioRxiv, and arXiv; since its release, it has served as the basis of many COVID-19 text mining and discovery systems [12]. The final release of June 2, 2022, indexes >1 million publications. As summarized in Figure 1, approximately 79% of the documents in CORD-19 have an abstract. Out of them, around 41% have a full-text JSON file available, while <11% of available full-text publications have no abstract in the metadata table. Thus, we decided to focus on data set records with an abstract. The file containing the metadata of publications in the data set is a comma-separated table (CORD-19 metadata.csv) including the following:

**Figure 1.** Euler-Venn diagram of the overlap of publications with abstract and publications with full-text JSON from PDF or from PubMedCentral (PMC) in the COVID-19 Open Research Dataset (CORD-19).

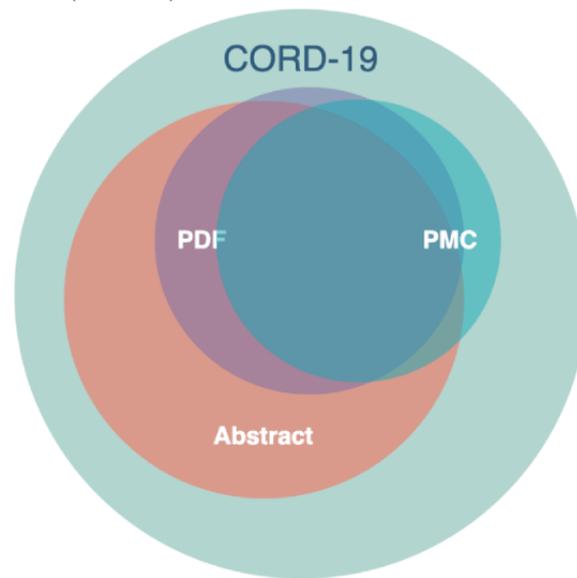

- A unique identifier cord_uid for a cluster of different records of the same publication—upon it, we performed deduplication and subsequent reconciliation of the other metadata of the cluster into a single record.
- Title of the publication—we detected the language and filtered out those not in English.
- Abstract of the publication—only records with an actual abstract were retained.
- Publish_time—the distribution of publication times, shown in Figure 2, shows that COVID-19 publications increased in the first half of 2020. Spikes at the beginning of each year correspond to subsequent publications whose publish time is incomplete (ie, only the year field was filled). Publications before 2020 that are concerned with Middle East

Respiratory Syndrome, Severe Acute Respiratory Syndrome, and the coronavirus were removed.
- Journal's abbreviated name—fuzzy matching of the abbreviated names was performed with a list of full names obtained from Scopus [13].
- Authors and DOI of the publication
- Number of citations received (ie, through numCitedBy), obtained by SemanticScholar application programming interfaces (APIs) [14]

Records from CORD-19 are already harmonized (refer to the study by Wang et al [12]), resulting in distinct cord_uid keys. However, several records of the same publication are included, with different metadata. We deduplicated them and retained just 1 record (the one published in a peer-reviewed journal, if available, else the richest one in metadata).

**Figure 2.** Line plot showing the 10-base logarithm of the number of publications (y-axis) per publish time and date (x-axis).

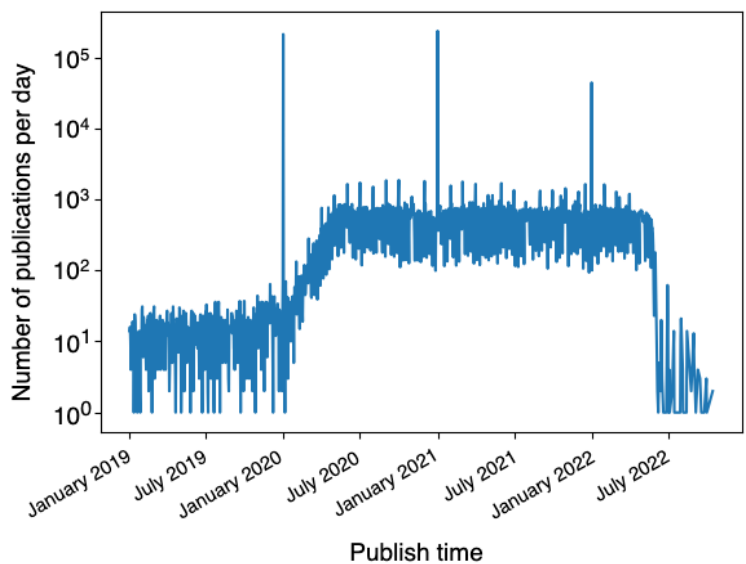

## Co-Occurrence Network

The co-occurrence network was built to support graph search; it consists of entities and relationships mined from the title and abstract fields of the metadata table. For building it, we considered 2 sources: UMLS and CIDO. UMLS [7] is a generic source that includes many vocabularies and covers the entire spectrum of medicine; CIDO [8] is a community-driven open-source biomedical ontology in the area of COVID-19.

While CIDO has a simple concept structure, UMLS concepts have a taxonomy that includes macrocategories at a coarse level; each macrocategory is further characterized by a type. Currently, we consider the following UMLS macrocategories: ACTIVITIES_AND_BEHAVIORS, ANATOMY, CHEMICALS_AND_DRUGS, CONCEPTS_AND_IDEAS, DEVICES, DISORDERS, ENTITY, GENES_AND_MOLECOLAR_SEQUENCES, GEOGRAPHIC_AREAS, LIVING_BEINGS, OBJECTS, OCCUPATIONS, ORGANIZATIONS, PHENOMENA, PHYSIOLOGY, and PROCEDURES.

As attributes, entities of the co-occurrence network include the name, an Umls_id when the entity is extracted from UMLS, and the frequency associated with the entity (ie, the number of documents in CORD-19 capturing that concept). Relationships in the co-occurrence network express the co-occurrence of 2 entities in ≥1 documents of CORD-19. Each relationship has the following attributes: a name (ie, built as concatenation in alphabetic order of the names of the entities that co-occur); a frequency (ie, the number of abstracts that mention such co-occurring entities); and several statistical indicators of the relationship's strength within the corpus, such as the pointwise mutual information value (comparing the relative frequency of 2 concepts occurring together in the text to the probability of either concept occurring independently [15]), the normalized pointwise mutual information (NPMI) value (normalized by the Shannon self-information, ranging from −1 to 1 [16]), and the Cramer V value (measuring the statistical significance of the co-occurrence between 2 entities [17]).

Figure 3 illustrates the process of ontology creation at a conceptual level. The process applies to textual abstracts (refer to Figure 3 where we consider an excerpt of the textual abstract of the study by Logette et al [18]) and consists of an entity recognition task aiming to extract the known ontological terms (ie, either from UMLS or from CIDO), followed by an entity linking task; eventually, we produce a co-occurrence network, whose entities are extracted terms and whose relationships connect entities that co-occur, weighted by the strength of the co-occurrence. Next, we detail the data extraction and transformation process.

Figure 3. Rationale of co-occurrence network construction. Ontological terms are recognized in textual abstracts using entity recognition; then, this process is reiterated with approximately 660,000 publications' abstracts. Terms are connected to each other using entity linking; each relationship between entities is associated with several properties representing the co-occurrence weight, using different statistical methods. The generated connected co-occurrence network has approximately 128,000 concepts and approximately 47 million relationships. ACE2: angiotensin-converting enzyme 2; CIDO: Coronavirus Infectious Disease Ontology; NPMI: normalized pointwise mutual information; UMLS: Unified Medical Language System.

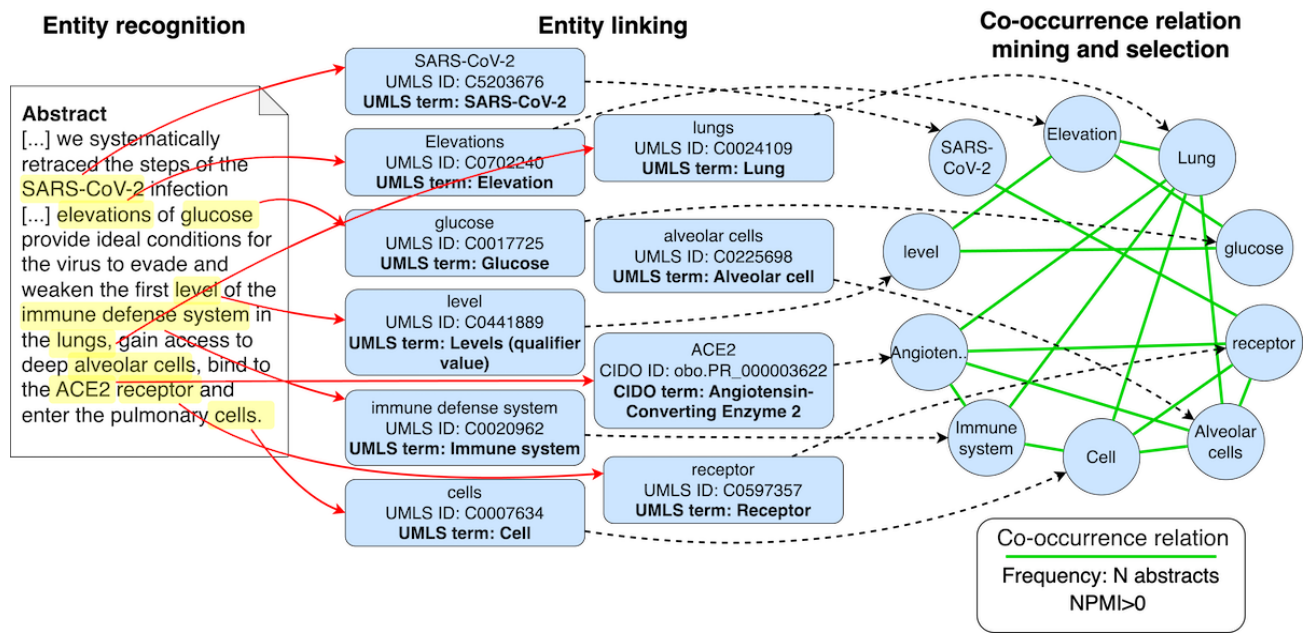

## Data Provisioning and Co-Occurrence Network Construction

### Overview

The data provision workflow is represented in Figure 4 [18]; it follows the extract-load-transform paradigm. Data were extracted from CORD-19 and loaded into the data storage system. The pipeline produces 3 data objects: the co-occurrence network; the metadata table; and the *inverted index*, that is, a simple postings list whose keys are the relationships of the co-occurrence network and whose elements are links to the relevant publications where such relationships co-occur. Other data tables contain intermediate results of the extraction and curation of the entities, that is, the nodes of the co-occurrence network and the computation of the co-occurrence measures used for the relationships. For storing data tables, we selected the MariaDB relational engine [19]; for storing the [20].
co-occurrence network, we selected the Neo4j graph data engine

**Figure 4.** Workflow diagram of the GRAPH-SEARCH data provision pipeline. Tasks are performed sequentially; each task uses data objects and produces data objects, starting from the raw COVID-19 Open Research Dataset (CORD-19) metadata.csv file present in CORD-19, which is translated into metadata.csv once cleaned. The final outcome of the pipeline is a Neo4j database containing the network.

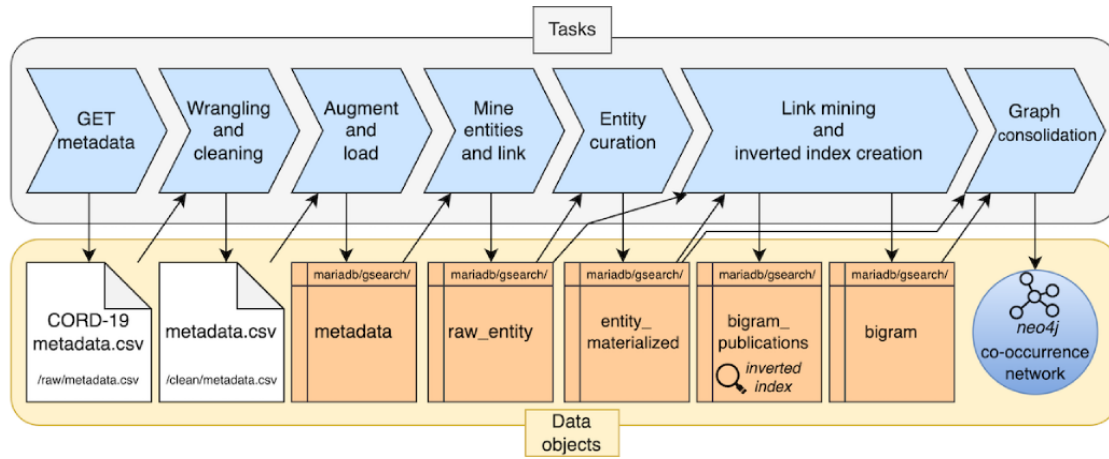

### Data Loading

Three tasks apply to raw CORD-19 data and produce a metadata table. Metadata were obtained by using the "GET metadata" from the S3 bucket of Allen Institute of AI; then, we performed a "Wrangling and cleaning" step and the "Augment and load" step on the cleaned metadata table with information from the external APIs.

### Entities Mining and Linking

The "Mine entities and link" task takes as input the curated and augmented metadata table and produces the raw_entity table. With a single pass over the title and abstract, we performed typical information retrieval steps such as lexical analysis, removal of stopwords, stemming, and lemmatization. Then, we performed named entity recognition (NER), consisting of the identification and extraction of entities from unstructured text and linking to UMLS and CIDO; specifically, we used the en_core_sci_lg model of the *scispaCy* Python package. The selected model is particularly suited for processing English-based scientific literature, providing an approximately 785,000 word vocabulary with 600k word vectors, with a declared $F_1$-score for mentions of 68.67 (refer to the study by Nuemann et al [21] for details on the achieved performances). Entities are linked to UMLS and CIDO by associating each concept with the UMLS identifier (with its related type and macroclass) and the CIDO identifier (if available).

### Entity Curation

The "Entity curation" task aggregates the occurrences in the raw_entity table and outputs the entity_materialized table, collecting all the entities to be used as nodes of the co-occurrence network. In this pass, we excluded the occurrences of the entities that score a low similarity with UMLS or CIDO concepts; we used a normalized string similarity measure based on the Levenshtein distance and a threshold value of 0.7. We also included within entities some *utility* terms that indicate level modifiers (eg, "high" and "increased") or causative connectors (ie, "induces"). Eventually, we added the entity type and macrocategory using their names in UMLS.

### Link Mining

The "Link mining and inverted index creation" task uses the raw_entity table and the entity_materialized table to generate the bigram table (ie, information on the links of the co-occurrence network) and the bigram_publications table that we use as an *inverted index* in the information retrieval process.

A co-occurrence is a relationship between 2 concepts, and it exists when those 2 concepts occur in the same document. Each relationship is named using the convention "X.name—Y.name," where X and Y are the 2 concepts expressed as nodes, which it connects, and X.name precedes Y.name alphabetically.

We designed a greedy algorithm—optimized for big data contexts—to extract the relationships in a single pass over the publications. This algorithm requires 2 *read-only* lookup tables, built before the execution: publication_entities (ie, for each publication a list of mentioned entities) and entity_publications (ie, for each entity, a list of mentioning publications). The complexity of the algorithm is $o(N^2)$, where $N$ is the number of entities in the entity_materialized list; in practice, the number of required comparisons is low, as the number of entities in each publication is much lower than the total number of entities selected in the "Entity Curation" step.

### Graph Consolidation

The "Graph consolidation" task selects data from the entity_materialized and bigram tables and migrates them to the Neo4j instance to create the co-occurrence network.

The nodes are curated in the previous "Entity Curation" step. The relationships of co-occurrence are chosen at this stage, based on their NPMI, which is the point estimate of the Mutual Information, normalized by the Shannon self-information (taking a value between −1 and +1); this compares the probability that the 2 entities occur together. We exclude the relationships with NPMI≤0, as a nonpositive NPMI indicates that the relationship is not significant.

The resulting co-occurrence network has 128,249 entities and 47,198,965 relationships, extracted from 662,105 initial publications. Using the Neo4j Graph Data Science library [22], we verified that the graph is a unique connected component; such a condition is essential to ensure that every possible formulated graph query can be matched on the co-occurrence network.

### Graph Query Search

A graph query $Q$ is a connected graph formed by nodes and undirected relationships, where nodes are the set of entities appearing in $Q$ and $rels(Q)$ is a set of arbitrary relationships connecting some pairs of entities in $Q$. A subgraph $Q'$ is simply a connected subset of the nodes and relationships of $Q$. The search strategy is composed of 2 steps: matching of graph query against the co-occurrence network and extracting the relevant publications.

**Figure 5.** Graph query matching operation.

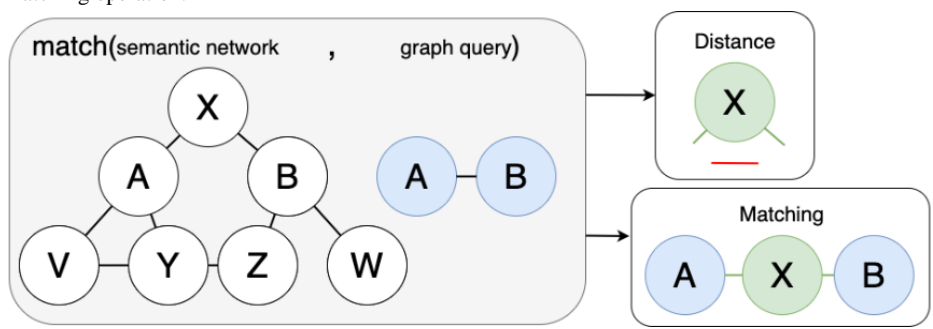

All entities in $Q$ are matched in $N$; then, for each relationship $r$ in $rels(Q)$, connecting nodes $\alpha$ and $\beta$, we retrieve the "shortest paths" within $N$ that connect $\alpha$ and $\beta$, that is, a chain of relationships $r_1', r_2', ..., r_n'$, where $r'$ is in $rels(N)$, $r_1'$ starts from node $\alpha$, and $r_n'$ ends in node $\beta$.

Shortest paths are computed using the All Pairs Shortest Path function allShortestPaths available in Cypher, Neo4j v4.4 [20]. Candidate shortest paths are ranked by the average of the NPMI property associated with each relationship along the path; we retain the top 10 paths in the ranking.

We refer to the set of candidate shortest paths as *expansion*; the selection of exactly 1 preferred path among the candidates of the expansion is performed interactively by the user of the search system, as it is strictly domain or context specific.

*Relevant publications extraction* corresponds to the retrieval of the publications that mention concepts of the matched graph, using the inverted index. We access the inverted index by relationship name, using either $r$ when it appears in the relationships $rels(N)$ of the co-occurrence network or all the relationships $r_1', r_2', ..., r_n'$ appearing in the specified $path(r)$. The score of a publication $P$ relative to a query $Q$ (ie, the number of explained relationships) is computed as follows:

$$Score(P, Q) = \sum_{r \in rels(Q)} \frac{\sum_{r' \in path(r)} P_{r'}}{|path(r)|}$$

*Graph query matching* is the operation of comparing the graph query $Q$ with the co-occurrence network $N$ created along the procedure described in the *Data Provisioning and Co-Occurrence Network Construction* section. By construction, each entity in $Q$ is contained in $N$, whereas relationships in $rels(Q)$, arbitrarily created in $Q$, may not be present in $N$. Both $Q$ and $N$ are connected graphs with undirected relationships; then, matching $Q$ within $N$ can be seen as an instance of inexact graph matching [23].

Figure 5 guides the intuition of the matching operation. A graph query A-B (in blue) is searched over a co-occurrence network (in white). No direct relationship exists between A and B on the network. However, several alternative finite paths exist (ie, A-X-B, A-Y-Z-B, or A-V-Y-Z-B). Among these, A-X-B is found to be the "shortest path" between A and B, as its length or distance (in green) equals 1.

The addends of the external summation represent a score assigned to each relationship $r$ in $Q$. Each addend captures how well $P$ represents $r$; it is equal to 1 if $P$ directly mentions the relationship of $Q$ (eg, when $path(r)=r'$, with length 1) or if $P$ mentions all the relationships of $path(r)$. Otherwise, it equals a fraction of 1, counting the number of relationships $r_1', r_2', ..., r_n'$ of $path(r)$ mentioned in $P$, divided by the length of $path(r)$.

Extracted publications are ordered by their score; they are further described by other properties, such as the sum of the NPMI of all the mentioned relationships and the date of publication.

### Running Example

Consider Figure 6 as an example of the four steps performed during the search:

1. *Create graph query* (Figure 6A): Nodes are chosen among the concepts existing in the co-occurrence network; node names can be found through a dedicated browser working either by autocompletion of user-typed content (ie, matching terminologies concepts) or by selection of category and type and the contained concepts; search on multiple terminologies at the same time is allowed. For each concept, we provide a description and ID from the original source. Relationships can be drawn between any pair of nodes.
2. *Find paths* (Figure 6B): For each pair of entities connected by a relationship in the graph query, the Neo4j graph is queried to find the shortest paths (at most 10) with top average NPMI scores.
3. *Select paths* (Figure 6C): The user selects the most relevant path for each original relationship that has been expanded.

4. *Retrieve publications and return ranking to the user* (Figure 6D): The system collects the names of all the relationships from the expanded graph query (computed in step B and selected in step C) and exploits them to retrieve the posting lists of publications (from the inverted index). It computes the relationships explained by each publication. Then, it ranks the publications by (1) the number of explained relationships of the original graph query (refer to equation 1), (2) the sum of NPMI scores of the relationships, and (3) the publication date. Finally, it shows the complete list with the publications' metadata.

**Figure 6.** (A) Example of a graph query with 6 concepts and 5 relationships. (B) Match of graph query on the co-occurrence network, with the search of shortest paths (in the dashed spaces called expansions). Considering the relationship between SARS-CoV-2 and angiotensin-converting enzyme 2, its expansion includes 3 paths of length 3, each characterized by 2 intermediate nodes. Light green paths have the highest average normalized pointwise mutual information (NPMI) of each expansion. (C) Regardless of the suggested paths with the highest average NPMI, users can select any path (dark green). (D) A list of publications, ranked by their score, is extracted; the score is computed using equation 1 and considers all the relationships in the selected paths that are mentioned in the publication. ACE2: angiotensin-converting enzyme 2; AngII: angiotensin II.

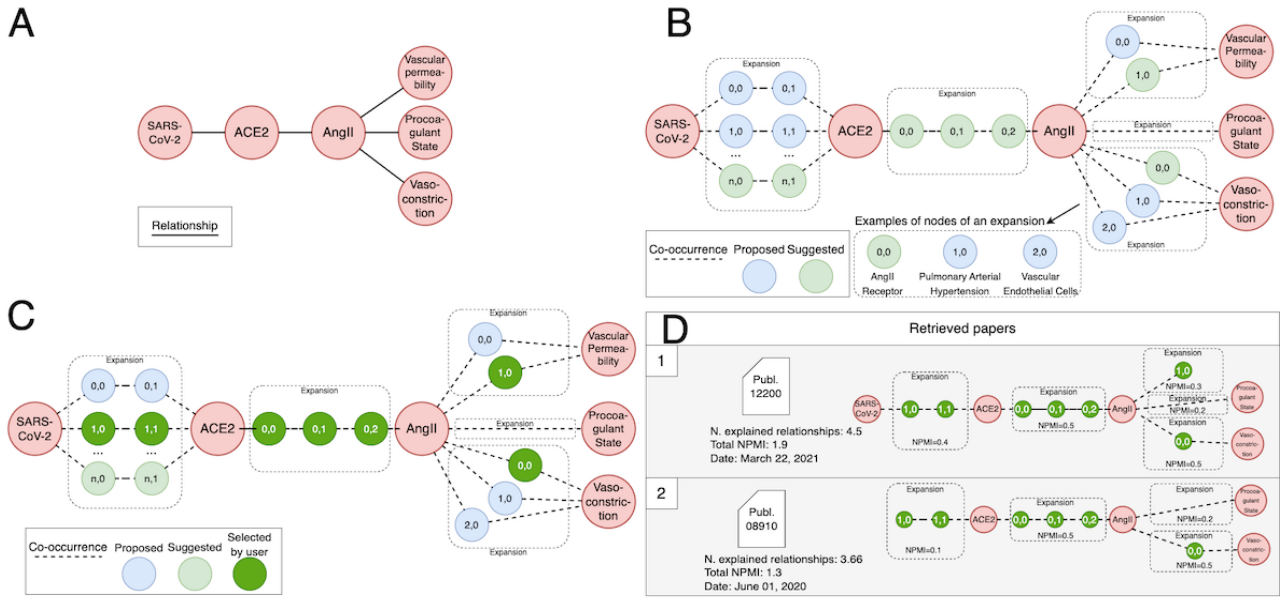

In Figure 6D, we observe that 5 expansions are produced: the first publication scores 1 in 4 expansions and 1/2 in the expansion at the top-right end of the graph query. Indeed, publication 1 only includes the relationship (AngII)-(1,0), which is half of the selected shortest path that connects (AngII) and (Vascular Permeability).

The second publication scores 0 in 1 expansion, as there is no path between (AngII) and (Vascular Permeability); 1 in 3 expansions; and 2/3 in the expansion at the left end of the graph query; the relationship (SARS-CoV-2)-(1,0) is not mentioned.

### Ethical Considerations

Ethics approval was not applicable for this study.

## Results

### Web Interface

With GRAPH-SEARCH, the researcher can express a query in the form of a graph query on a web interface and retrieve a list of CORD-19 publications that best correspond to the query. During the search process, each link in the original graph query is expanded and matched with the co-occurrence network. When a relationship in the query is not available in the co-occurrence network, an expansion may suggest that several sets of concepts can explain a relationship in the original graph query; therefore, 10 ranked paths are proposed to the user, who may express a preference according to their interest. After selecting 1 path for each expanded relationship, GRAPH-SEARCH provides a list of publications ranked by the number of explained relationships of the original graph query.

The GRAPH-SEARCH application service exposes a web user interface to query the co-occurrence network and exploit the graph-driven search methodology described in the *Graph Query Search* section; it contains a backend (ie, web server that exposes a Representational State Transfer Application Programming Interface for high-level retrieval operations) and a frontend (ie, visual interface that exploits the Representational State Transfer Application Programming Interfaces to use the backend).

The web interface has been designed and implemented following the major steps of the algorithm described in the *Running Example* subsection above. The user experience has been modeled as a multipage application; for each step of the retrieval strategy, different API services and a different page were implemented.

The frontend is built with the Vue.js framework and the D3.js library for graph illustrations; instead, the backend is written in Python and includes two components:

1. Swagger_server, which implements the web service logic, interfaces, and the models necessary to handle the persistence and asynchronicity behaviors of a multiuser system. We used the connexion framework, a flask-based web framework, and SQLAlchemy as the database abstraction layer.

2. Core, which implements the retrieval strategy and provides high-level programming interfaces for it. This package has been designed as an independent library that can be embedded in other applications, as it has been done with the backend service. Its implementation relies on several Python libraries, such as Neo4j, networkx, and SQLAlchemy.

## UC Queries

UC1 emphasizes the strength of exploratory search over graphs by supporting users in selecting graph portions, considering eventually accepting proposed expansions, and browsing results in terms of NPMI and explained relationships. UCs of increasing complexity are provided next, offering examples of searches upon graph queries with different shapes: UC2 and UC3 introduce very simple linear graph queries (ie, 1 chain of nodes), UC4 shows the use of a Y-shaped graph query, and UC5 and UC6 represent more complex shapes with nodes forming triangles.

## UC1: Genetic Mechanisms of Critical Illness in COVID-19

Pairo-Castineira et al [23] revealed previously undescribed molecular mechanisms of critical illness in patients with

COVID-19 with genome-wide studies. The results of such studies may provide therapeutic targets to modulate the host immune response to promote survival. Inspired by this publication, we create a graph query including relevant human genes that are related to higher or lower severity of COVID-19 (eg, *IFNAR2, CCR2, and TYK2* genes), and we link them to the change in the severity of the disease (Figure 7A). Since the research idea is broad, we start the exploratory process focusing on a subgraph of the graph query (refer to the nodes in red selected in Figure 7A). Here, we only consider the effect of the increase of expression in the *CCR2* gene. Figure 7B shows how GRAPH-SEARCH expands the path between the concepts "High" and "Gene Expression" (not otherwise connected in the co-occurrence network). According to NPMI values, the most relevant concept connecting them is "Up-Regulation (Physiology)." Figure 7C shows that the path going through this concept has been selected by the user among the other proposed. The Results page (Figure 7D) shows a publication (Teixeira et al [24]) that covers 4 (80%) out of 5 explained relationships of the original graph query. This means that out of the 5 original relationships of the selected portion of the graph query, only 4 (80%) are explained by the publication (all except for the one between "Gene Expression" and "High"). At this point, the user can consider other portions of the graph query or the entire query.

**Figure 7.** GRAPH-SEARCH screens dedicated to use case 1 (UC1): (A) graph query, (B) find paths, (C) select paths, and (D) first publication on the results page.

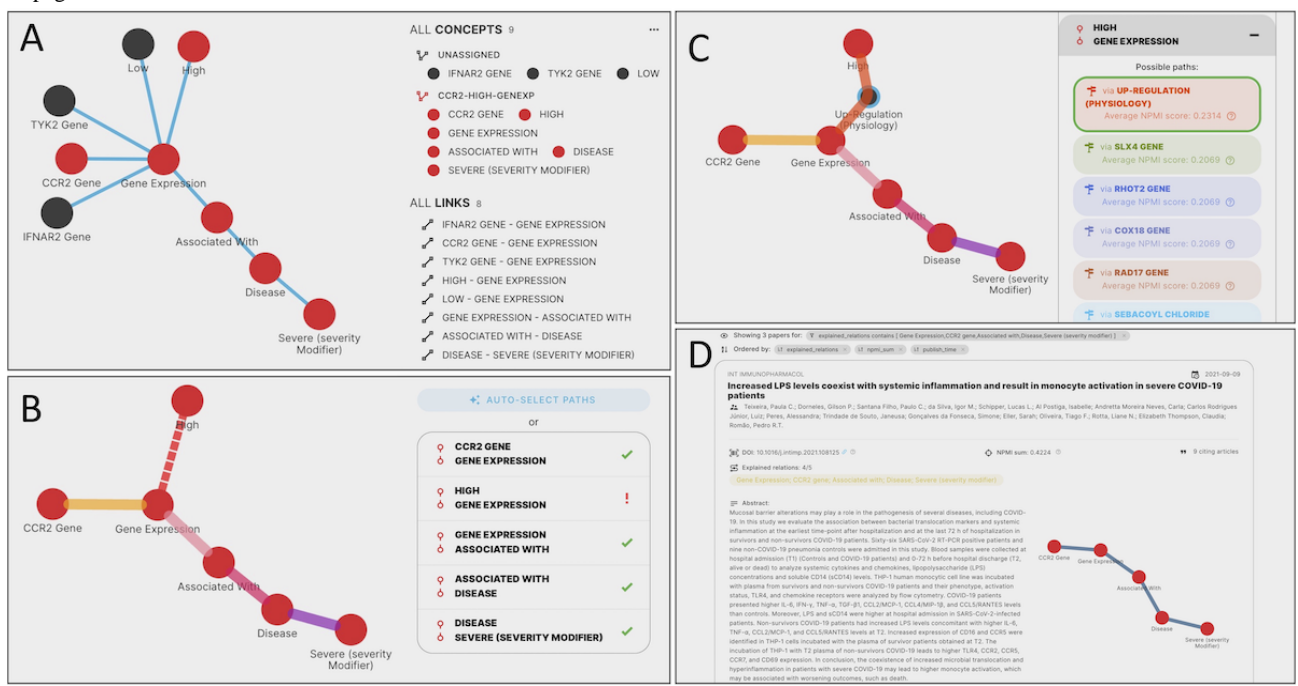

## UC2: COVID-19 and Cystic Fibrosis

Cystic fibrosis is a disorder that affects mostly the lungs, the digestive system, and other organs in the body. It is widely known that COVID-19 also affects the respiratory system. *How has their connection been investigated in CORD-19?* The simplest possible graph query in GRAPH-SEARCH holds 2 nodes (ie, cystic fibrosis and COVID-19) connected by 1 relationship of co-occurrence. "Cystic fibrosis" is represented by UMLS concept ID C0010674, and "COVID-19" is represented by the UMLS concept ID C5203670. The 2 concepts are not directly connected within the network; among the proposed paths in the expansion, we choose the one through the concept "Respiratory secretion viscosity alteration" (UMLS ID 3537094). Only 1 publication in CORD-19 explains this path, covering it completely, with an NPMI sum of 0.5668. Kratochvil et al [25] characterized the composition of respiratory secretions of intubated patients with COVID-19 infection, finding that they closely resemble those of cystic fibrosis, a minor observation unrelated to clinical severity. In general, the lack of relevant clinical references confirmed our expectation that cystic fibrosis did not impact COVID-19 severity.

### UC3. COVID-19 and Nonsteroidal Anti-Inflammatory Drugs

During the second year of the pandemic, interest arose in the possibility of intervening at the onset of mild to moderate COVID-19 symptoms in outpatients (instead of hospitalized patients); it was suggested that this could prevent the progression to a more severe illness and long-term complications. More specifically, Perico et al [26] investigated the use of anti-inflammatory drugs, especially nonsteroidal anti-inflammatory drugs (NSAIDs) as a therapeutic strategy. In our graph query, we include the following as main concepts: "COVID-19" (C5203670), "Outpatients" (C0029921), "Anti-Inflammatory Agents, Non Steroidal" (C0003211), and "Cyclooxygenase 2 Inhibitors" (C1257954), with the last being a specific class of NSAIDs. In this case, no expansion of the original graph query is performed, as all the relationships are present in the co-occurrence network. The Results page contains a list of 440 publications, whose abstracts discuss the concepts in the graph query from different perspectives and approaches. The top 3 results include work from Consolaro et al [27], a home-treatment algorithm based on anti-inflammatory drugs; Popovych et al [28], discussing the therapeutic efficacy of the BNO 1030 extract, which is a phytotherapeutic anti-inflammatory agent; and Sava et al [29], exposing the results of a 90-day treatment of patients with severe COVID-19 with a specific NSAID drug, tocilizumab.

### UC4: Elevated Blood Glucose Levels and COVID-19 Severity

Elevated blood glucose levels are considered a risk factor for the severity of the disease. With GRAPH-SEARCH, we compose a Y-shaped graph query (Figure 8), expressing that high levels of blood glucose or increasing blood glucose can induce a severe illness. This example makes sophisticated use of *utility* terms; these are provided in a specific list of the concepts' browser of GRAPH-SEARCH.

**Figure 8.** Graph query of use case 4 (UC4), with Unified Medical Language System concepts IDs in red.

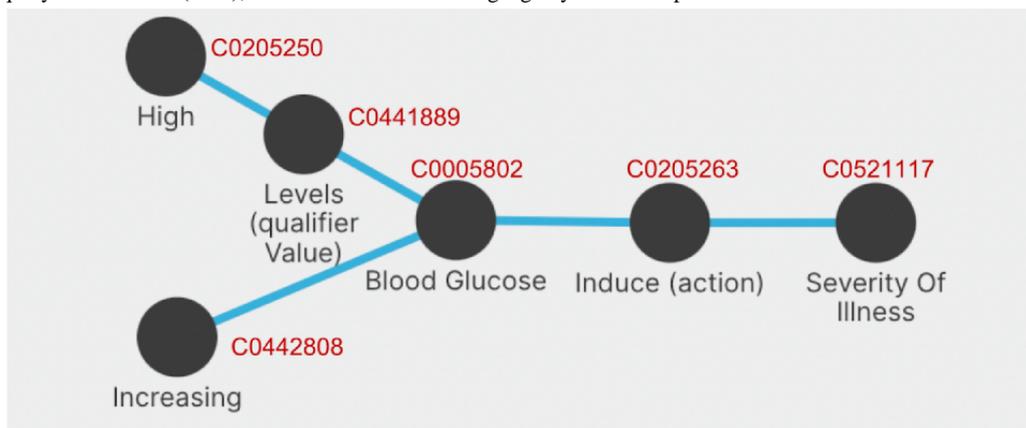

Consequently, we obtain a list of 395 results, where the top-ranked publication explains 5 out of 5 relationships. Logette et al [18] reported on the relationship between blood glucose levels and the severity of COVID-19. All following publications, ranked in descending order by the number of explained relationships of the original graph query, explain at most 3 out of 5 relations.

### UC5: COVID-19, Angiotensin-Converting Enzyme 2, and Cardiovascular Diseases

Patel et al [30] hypothesized that the infection caused by SARS-CoV-2 could be associated with the shedding of angiotensin-converting enzyme 2 (ACE2). In their study, it is suggested that in patients with cardiovascular diseases, there is increased shedding of ACE2; consequently, higher levels of ACE2 in blood circulation are associated with the downregulation of membrane-bound ACE2. The graph query in Figure 9A expresses this query by connecting patients with COVID-19 infection with cardiovascular diseases; as they have more circulating ACE2, there is a downregulation of membrane-bound ACE2. When running this query, 2 relationships are not found in the co-occurrence network; the first paths suggested by the system as possible explanations are not meaningful with regard to the context; thus, we select alternative concepts, that is, "Subacute Endocarditis" and "Intensive Care Unit" (Figure 9B). Results can be ranked by the number of citations; we found the studies by Yamaguchi et al [31] and Gupta et al [32] particularly interesting, as they propose solutions for the prevention and treatment of the side effects of COVID-19 for patients with cardiovascular diseases.

**Figure 9.** (A) Graph query of use case 5 (UC5) and (B) found paths.

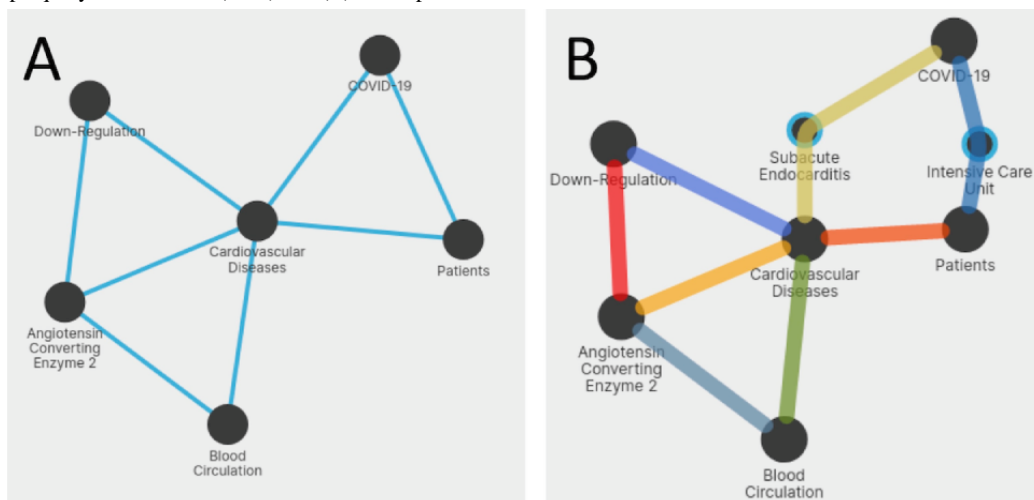

### UC6: COVID-19 Vaccines and Myocarditis

The side effects of vaccines are a topic of relevance. Here, we investigate the connection between events of heart inflammation (eg, myocarditis) among adolescents and the COVID-19 Moderna vaccine. We compose a graph query in GRAPH-SEARCH with 4 nodes (Figure 10A); a triangle is formed by "Adolescent (age group)" (C0205653), "Myocarditis" (C0027059), and the "Moderna COVID-19 Vaccine" (CIDO ID obo.VO 0005157); the vaccine entity is connected to the "COVID-19" (C5203670) node. COVID-19 and Moderna COVID-19 vaccine are not directly connected; among the possible paths suggested by GRAPH-SEARCH, the 2 scoring the highest sum of mutual information are through "Vaccination" and "Myopericarditis." The latter refers to both myocarditis and pericarditis (ie, the inflammation of the pericardium, which is the sac that surrounds the heart). The latter concept allows us to expand the initial query to complete the match with the co-occurrence network (Figure 10B). On the Results page, 190 bibliographic resources are provided. The top-ranked one, which explains all 4 relationships of the graph query, is a report by Gargano et al [33] that highlights the implications of the use of messenger RNA vaccines with a higher risk for myocarditis in male individuals aged 12 to 29 years. The following results do not explain the relationship between the COVID-19 Moderna vaccine and COVID-19 through myopericarditis, as they explain only 3 relations. These results, for instance, report adverse events of myocarditis after vaccination in the United States [34] and Korea [35].

**Figure 10.** (A) Graph query of use case 6 (UC6) and (B) found paths.

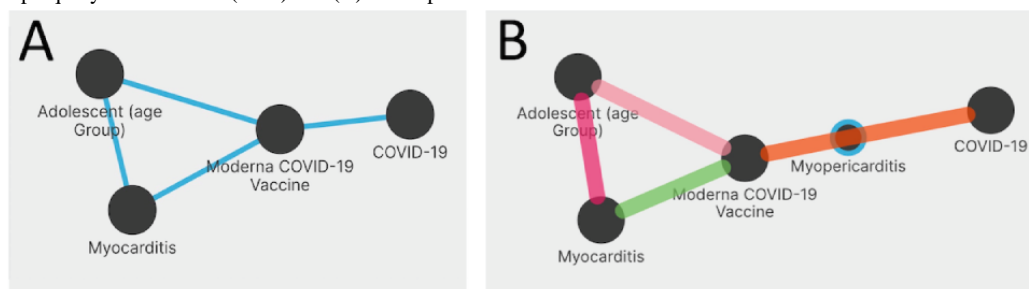

### Query Performances

GRAPH-SEARCH queries are composed of two computationally intensive steps: (1) the graph query matching over the co-occurrence network and (2) the retrieval and ranking of publications related to the query. For each such step, we run a performance analysis.

Specifically, we simulated random queries with 2, 4, 6, 8, or 10 nodes from the existing co-occurrence network; we assume that these are the typical UC scenarios, as queries represent small queries of researchers created through the graphical interface.

We separately measure computation times of the first and second steps (shown in Figures 11A and 11B, respectively); each experiment has been repeated on 10 queries, generated randomly using the "Random walk with restarts sampling" method of Neo4j. We observe that the computational times for graph matching in all cases is <2.2 seconds, and its growth is less-than-linear with the number of the nodes, whereas the retrieval operation typically takes up to 3 seconds, with a small number of outliers due to cache misses; the resulting user delay in these scenarios seems quite acceptable.

**Figure 11.** Box plots measuring the time for (A) the graph matching operation and (B) the publication retrieval operation performed using complete graph queries of 2, 4, 6, 8, and 10 nodes (each repeated 10 times).

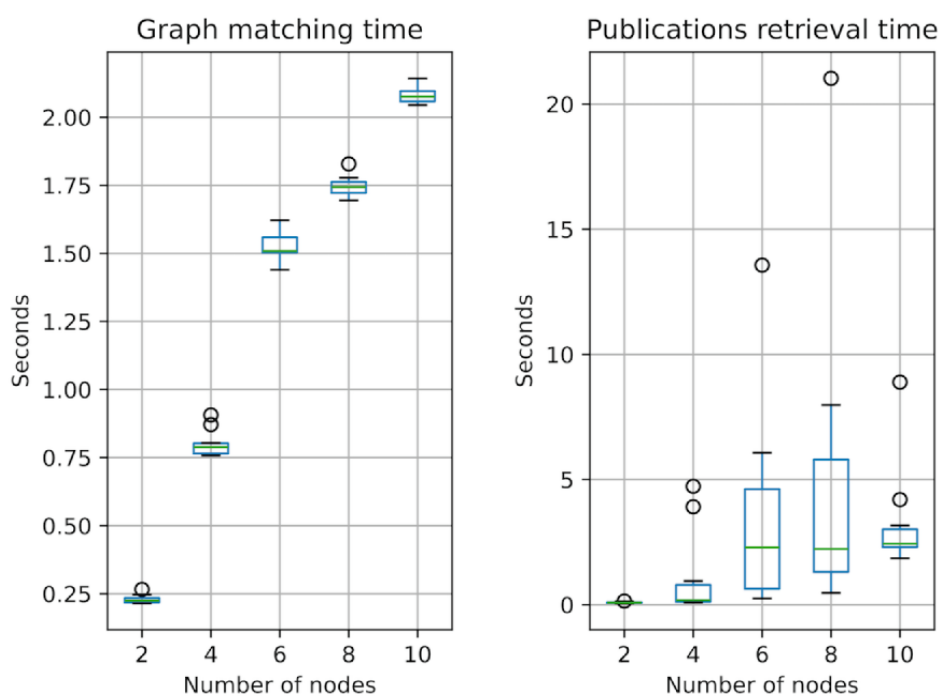

We also created random graph queries by removing increasing percentages of their relationships to simulate the difference between exact and inexact graph search (thereby triggering the search for alternative shortest paths); computational times (not shown for brevity) are not significantly affected.

## Related Work

In this section, we review classic approaches to search over co-occurrence networks. Then, we focus on the specific use of bio-ontologies in information extraction systems, and finally, we propose a close comparison with COVID-19–specific search systems.

### Semantic-Network Search

The task of searching and extracting literature documents over co-occurrence networks with graph-based queries can be considered through the subproblems that compose it. To query a co-occurrence network with a graph-like query, a similarity measure between graphs must be defined. Existing methods in the context of graph databases include definitions of graph edit distances and maximum common subgraphs [36], but a later approach introduced a similarity measure based on a graph kernel between pairs of documents, which exploits the shortest paths between nodes as units to compare graphs [37]. Considering the construction of the co-occurrence networks from data sets of literature documents, different approaches are available to extract concepts to represent nodes in the network and connections between them. The survey by Han et al [38] and the study by Shi et al [39] present all the main methodologies and text mining pipeline architectures, which are applied in this study to engineering and design (ie, subsets of scientific literature). G-Bean [40] is also a relevant related work, that is, a graph-based tool that exploits ontologies for graph-based query expansion to support the user search intention discovery.

### Literature Annotation and Bio-Ontologies

The incorporation of bio-ontologies in information extraction and information retrieval has demonstrated its efficacy through diverse applications, such as patent information retrieval [41] and identification of concept domains [42]. Bio-ontologies are also applied in natural language processing tasks, such as NER [43]. Moreover, Wang et al [44] illustrated the application of bio-ontologies in retrieving biomedical data sets, while Maraver et al [45] emphasized their role in literature search facilitation and metadata organization. The potential for refining search queries through ontology-guided expansion is also a recurring theme in the biomedical literature for information retrieval. Diaz-Galiano et al [46] and Dong et al [47] show query expansion methodologies using different medical vocabularies.

A fundamental aspect of research in this domain pertains to the availability and use of suitable corpora and data sets; previous studies [48,49] have provided foundational annotated and curated resources that underpin the experimental frameworks addressing these tasks. Lately, the integration of bio-ontologies with language models has also gained traction within the context of bioinformation extraction [50,51].

### COVID-19–Specific Literature Discovery

With the outbreak of the COVID-19 pandemic, several open-access data sets have been collected, including the National Institute of Health's COVID-19 [52], the Human Coronaviruses Data Initiative [53], and COVIDScholar [54].

The CORD-19 data set received the widest attention. Several knowledge graphs that exploit this data set were proposed at the beginning of the pandemic for representing biomedical entities (eg, CORD-NER [55] and COVID-19 KG [56]) or publications metadata (eg, COVID-19-Literature [57]). More recently, CovidPubGraph [58] has provided a comprehensive and updated knowledge graph, which integrates information from multiple sources, making results available through a SPARQL end point. Finally, CovidGraph [59] exposed a knowledge graph in the Neo4j browser; several external ontologies are used to tag entities. The focus of these resources is more on organization and semantic enrichment than on exploration.

The goal of the TREC-COVID initiative [60] was to establish targeted retrieval tasks in response to the pandemic, to be shared and collectively addressed by the community. Instead, GRAPH-SEARCH aims to make the literature about COVID-19 searchable and explorable. This objective is common to other 2 systems, LitCovid and Outbreak.info; these support enhanced keyword-based search, but they do not offer any graph-based search support.

LitCovid [1] was developed within the US National Institutes of Health as a comprehensive resource of literature on COVID-19 (372,221 publications at the time of writing), updated regularly starting from PubMed. Publications are manually screened to assess their relevance to COVID-19. They are then categorized (eg, overview, disease mechanism, transmission dynamics, treatment, case report, and epidemic forecasting); assigned geographical locations; and annotated with any drug or chemical-related information found in their title and abstract, if applicable. The updated version [61] introduced the long-covid category, added annotations on variants and vaccines, and supported with machine learning algorithms the topic categorization (with a more updated model) and entity recognition (with NER). The interface allows us to apply filters on country, journal, drug, variant, and vaccine and compose search strings combining AND, OR, and NOT operators (ie, not documented); results are ranked by relevance, based on the widely used BM25 ranking function of Lucene. LitCovid positively compares its performances to the classical keyword search of PubMed (where annotations or tags are not used).

Outbreak.info Research Library [2] is a project of the Hughes, Su, Wu, and Andersen laboratories at Scripps Research. It offers a searchable interface of COVID-19 publications (complementing the content of LitCovid integrating preprint servers), together with clinical trials, data sets, protocols, and other resources. The data structure upon which the search is performed is supported by a schema; entities are connected by links with various semantics. The visual interface allows the use of some filters and keyword search; results are ranked by relevance based on the Lucene Practical Scoring Function on Elasticsearch (prioritizing the query normalization factor, coordination factor, term frequency, and inverse document frequency).

## Discussion

In this section, we discuss how the proposed graph query search could be compared to other information extraction setups. For this purpose, we focus on 2 UC queries, that is, the linear query presented in UC3 (4 nodes in a linear pattern) and the red subgraph shown in UC1 (a nonlinear 6 nodes query, expanded with an additional node in GRAPH-SEARCH).

### Comparison With COVID-19 Literature Search Systems

First, we considered running the UCs on the COVID-19 literature–dedicated search systems LitCovid and Outbreak.info. Both systems were queried using concept names corresponding to UMLS terms in the nodes; unfortunately, they both suffer from the limitations of Boolean search. Specifically, if we search with conjunctive clauses and exact search (eg, using "Outpatients" AND "Anti-Inflammatory Agents, Non Steroidal" AND "Cyclooxygenase 2 Inhibitors" AND "COVID-19" for UC3), no system returns any result. Dealing with exact search is hard. For instance, with LitCovid, the query "Cyclooxygenase Inhibitors" produces 3 results, whereas the query "Cyclooxygenase 2 Inhibitors" produces 5 results, although apparently more restrictive; instead, the query Cyclooxygenase Inhibitors (no quotes), without exact search, produces 12,287 results (including all references referring to generic inhibitors). Table 1 reports the results of LitCovid with conjunctive queries but no exact matching, while a similar search is not supported by Outbreak.info. In comparison, GRAPH-SEARCH reports 327 results for UC1 and 440 results for UC3. These outputs are hardly comparable, mainly because with LitCovid it is not possible to build a unique graph-shaped query; therefore, results of single conjunctive queries need to be evaluated one after the other, whereas GRAPH-SEARCH aggregates together the results of several conjunctive chains; it also expands given concepts with their acronyms (eg, "anti-inflammatory agents, non steroidal" is also searched as "NSAIDs"). In addition, GRAPH-SEARCH allows for the expansion of specific links by adding new concepts (eg, "Up-Regulation [Physiology]" in UC1). No domain-specific system for COVID-19 supports graph-based search, allowing a more insightful comparison.

**Table 1.** Results of the evaluation of use case (UC) 1 (Figure 7) and UC3 queries when performed on the LitCovid search interface, on the full-text indexed MariaDB database, and on GRAPH-SEARCH.

| Query | Retrieved pubs LitCovid, n | Retrieved pubs MariaDB, n |
|---|---|---|
| **UC1: GRAPH-SEARCH retrieves 327 publications overall** | | |
| (Severe (severity modifier)) AND (Disease) AND (Associated With) AND (Gene Expression) AND (High) AND (CCR2 gene) | 316 | 11 |
| (Severe (severity modifier)) AND (Disease) AND (Associated With) AND (Gene Expression) AND (High) AND (CCR2 gene) AND Up-Regulation (Physiology) | 52 | 12 |
| (Outpatients) AND (Anti-Inflammatory Agents, Non Steroidal) AND (Cyclooxygenase 2 Inhibitors) | 972 | 4 |
| **UC 3: GRAPH-SEARCH retrieves 440 publications overall** | | |
| (Outpatients) AND (Anti-Inflammatory Agents, Non Steroidal) AND (COVID-19) | 1714 | 3 |
| (Outpatients) AND (Cyclooxygenase 2 Inhibitors) AND (COVID-19) | 3018 | 1 |
| (Anti-Inflammatory Agents, Non Steroidal) AND (Cyclooxygenase 2 Inhibitors) AND (COVID-19) | 37322 | 5 |
| (Outpatients) AND (Anti-Inflammatory Agents, Non Steroidal) AND (Cyclooxygenase 2 Inhibitors) AND (COVID-19) | 902 | 5 |
| "Outpatients" AND "Anti-Inflammatory Agents, Non Steroidal" AND "Cyclooxygenase 2 Inhibitors" AND "COVID-19" | 0 | 5 |

## Comparison With the Search on Full-Text Indexed Corpora

We also attempted a comparison with search operations performed on a baseline created by full-text indexing the CORD-19 titles and abstracts. Specifically, we used the full-text indexing option of MariaDB, an open-source fork of MySQL [19]. Typically, full-text indexes work well for regular text; they build an index over specific words rather than the whole text, and consequently, they show good performances for searches of specific words. The same queries used on LitCovid and Outbreak.info were used on this setup: on MariaDB, we used the "Natural language mode" documented on MariaDB [62] and, thus, we removed the "AND" Boolean operators and parentheses. To be part of the index, words must appear in <50% of the documents to be considered potentially relevant and to be used in searches (consequently, "COVID-19" and "SARS-CoV-2" are not considered relevant). Results are returned in descending order of relevance; limitations include the exclusion of partial (or very short or long) words.

Notwithstanding our attempts, we note that the comparison of the GRAPH-SEARCH approach with the full-text indexing setup is very difficult for many reasons:

1. The databases upon which search is performed are built on different assumptions (eg, to be part of the index, words must appear in <50% of the documents; the co-occurrence network only includes entities that score high similarity with ontology concepts and exclude relationships with a negative NPMI).
2. In 1 case, we perform separate keyword search sessions with separate results (with associated precision and recall measures); in the other case, we retrieve aggregated results (with summarized measures).

3. On one side, the ranking produced is only on single query result sets; on the other side, it is a global ranking.

The results are reported in Table 1; they must be read considering all these aspects. Note that results achieved with keyword search are restricted to manipulating Boolean expressions, adding, keywords and dropping keywords. On the contrary, the results on GRAPH-SEARCH (327 and 440, respectively for UC1 and UC3) are inspectable, with ranking, ordering, filtering, and visualization options dedicated to the explained chains of entities; using our search paradigm, users can compose graph queries; more complex topologies also allow a stronger explainability of results.

## Conclusions

GRAPH-SEARCH is the first search engine to propose the exploration of COVID-19 scientific literature using visual graph queries. GRAPH-SEARCH provides several unique features such as the possibility to describe concepts using well-known ontologies, to establish co-occurrence relationships between any 2 concepts of choice, to support search queries with concepts proposed and ranked by the system, and to browse resulting publications exploiting several visual and analytic measures.

The completeness and accuracy of the information captured in the co-occurrence network strictly depend on the advances of the NER methods used during the steps of entity mining and linking. Other systems have used expert curation (eg, LitCovid) or community-driven curation (eg, Outbreak.info). Although expert curation can improve the search experience, it does not properly scale; we opted for the exploitation of well-known biomedical ontologies such as UMLS and CIDO and state-of-the-art natural language processing models used for Entity Recognition in our data provision pipeline.

The ability of our system to extract results was evaluated, attempting a comparison with existing published systems (eg,

LitCovid and Outbreak.info) and with full-text indexing search. We recognize that comparisons between the results retrieved from these systems are not ideal, as it is very critical to compare single search runs with a system where the result is built progressively on the graph—considering a set of aspects altogether (ie, how the network was built and pruned, shortest path computation, completion with additional nodes, and global ranking of results).

Co-occurrence networks are conventionally used for analyzing extensive text and big data. Common applications have involved sentiment analysis [63] and detection of prevailing topics [64]. Here, each node is a word occurring in a set of user-generated social media posts. Moreover, word-co-occurrence networks are present in clinical applications, for example, Millington and Luz [65] proposed to encode recordings of speech data used for recognizing patients with Alzheimer and controls. In all such cases, GRAPH-SEARCH may be used to find specific subgraphs and propose completions of missing links.

In this study, we have demonstrated the capability of domain-specific (even inexact) graph query matching when semantics is considered only for nodes; we are aware of the limitations of this approach, which, at this stage, is considered a modeling choice. In future work, we plan to extend our search system to semantically rich knowledge graphs with both entities and relationships, thereby enriching the expressivity of graph queries (also including the possibility to capture the semantics of relationships, with state-of-the-art methods [66] or as we already experimented in a previous study [67]). Then, we aim to formalize the use of graph queries in the context of graph databases by studying the complexity of graph search and connecting it to classical theories of subgraph matching, shortest path search, and conjunctive query processing.

We also aim to conduct extensive empirical studies to measure user satisfaction with systems such as GRAPH-SEARCH analyzed along the 3 dimensions of usability, usefulness in deepening their knowledge of certain connected topics, and support of user's intentions in knowledge exploration.

## Acknowledgments

The authors would like to thank Luca Minotti for implementing the frontend of the GRAPH-SEARCH web application. This research is supported by the PNRR-PE-AI FAIR project funded by the NextGenerationEU program.

## Data Availability

The data processing pipeline is available as a Docker image [68]. The GRAPH-SEARCH application is available [69] and documented in the wiki [70].

## Authors' Contributions

AB and SC conceived the work; FI, AB, and SC jointly conceptualized and designed the framework; FI implemented the pipelines of data ingestion, network construction, and graph query engine; FI and AB curated the user experience; AB drafted the manuscript; FI and SC improved it; all authors revised the final version of the manuscript; and SC supervised the project.

## Conflicts of Interest

None declared.

## References

1. Chen Q, Allot A, Lu Z. LitCovid: an open database of COVID-19 literature. Nucleic Acids Res. Jan 08, 2021;49(D1):D1534-D1540. [FREE Full text] [doi: 10.1093/nar/gkaa952] [Medline: 33166392]
2. Tsueng G, Mullen JL, Alkuzweny M, Cano M, Rush B, Haag E, et al. Outbreak.info Research Library: a standardized, searchable platform to discover and explore COVID-19 resources. Nat Methods. Apr 2023;20(4):536-540. [FREE Full text] [doi: 10.1038/s41592-023-01770-w] [Medline: 36823331]
3. Kejriwal M. Knowledge graphs and COVID-19: opportunities, challenges, and implementation. Harvard Data Science Review. Dec 1, 2020. URL: https://hdsr.mitpress.mit.edu/pub/xl0yk6ux/release/4 [accessed 2024-05-08]
4. Peng J, Xu D, Lee R, Xu S, Zhou Y, Wang K. Expediting knowledge acquisition by a web framework for Knowledge Graph Exploration and Visualization (KGEV): case studies on COVID-19 and Human Phenotype Ontology. BMC Med Inform Decis Mak. Jun 02, 2022;22(Suppl 2):147. [FREE Full text] [doi: 10.1186/s12911-022-01848-z] [Medline: 35655307]
5. Ware C. Visual queries: the foundation of visual thinking. In: Tergan SO, Keller T, editors. Knowledge and Information Visualization. Berlin, Heidelberg. Springer; 2005.
6. Cheerkoot-Jalim S, Khedo KK. Literature-based discovery approaches for evidence-based healthcare: a systematic review. Health Technol (Berl). 2021;11(6):1205-1217. [FREE Full text] [doi: 10.1007/s12553-021-00605-y] [Medline: 34722102]
7. Bodenreider O. The Unified Medical Language System (UMLS): integrating biomedical terminology. Nucleic Acids Res. Jan 01, 2004;32(Database issue):D267-D270. [FREE Full text] [doi: 10.1093/nar/gkh061] [Medline: 14681409]
8. He Y, Yu H, Ong E, Wang Y, Liu Y, Huffman A, et al. CIDO, a community-based ontology for coronavirus disease knowledge and data integration, sharing, and analysis. Sci Data. Jun 12, 2020;7:181. [FREE Full text] [doi: 10.1038/s41597-020-0523-6] [Medline: 32533075]

9.  Rindflesch TC, Kilicoglu H, Fiszman M, Rosemblat G, Shin D. Semantic MEDLINE: an advanced information management application for biomedicine. Inf Serv Use. Sep 06, 2011;31(1-2):15-21. [doi: 10.3233/ISU-2011-0627]

10. Bernasconi A, Canakoglu A, Masseroli M, Ceri S. META-BASE: a novel architecture for large-scale genomic metadata integration. IEEE/ACM Trans Comput Biol Bioinf. 2022;19(1):543-557. [doi: 10.1109/tcbb.2020.2998954]

11. Fernandez JD, Lenzerini M, Masseroli M, Venco F, Ceri S. Ontology-based search of genomic metadata. IEEE/ACM Trans Comput Biol Bioinf. Mar 1, 2016;13(2):233-247. [doi: 10.1109/tcbb.2015.2495179]

12. Wang LL, Lo K, Chandrasekhar Y, Reas R, Yang J, Burdick D, et al. CORD-19: the COVID-19 Open Research Dataset. arXiv. Preprint posted online on April 22, 2020. [FREE Full text] [doi: 10.48550/arXiv.2004.10706]

13. Rose ME, Kitchin JR. pybliometrics: scriptable bibliometrics using a Python interface to Scopus. SoftwareX. Jul 2019;10:100263. [doi: 10.1016/j.softx.2019.100263]

14. danielnsilva / semanticscholar. GitHub. URL: https://github.com/danielnsilva/semanticscholar [accessed 2023-11-15]

15. Church KW, Hanks P. Word association norms, mutual information, and lexicography. In: Proceedings of the 27th Annual Meeting on Association for Computational Linguistics. 1989. Presented at: ACL '89; June 26-29, 1989; Vancouver, BC. [doi: 10.3115/981623.981633]

16. Bouma G. Normalized (pointwise) mutual information in collocation extraction. In: Proceedings of Conferences of the German Society for Computational Linguistics and Language Technology. 2009. Presented at: GSCL 2009; September 30, 2009; Potsdam, Germany.

17. Cramér H. Mathematical Methods of Statistics. Princeton, NJ. Princeton University Press; 1999.

18. Logette E, Lorin C, Favreau C, Oshurko E, Coggan JS, Casalegno F, et al. A machine-generated view of the role of blood glucose levels in the severity of COVID-19. Front Public Health. Jul 28, 2021;9:695139. [FREE Full text] [doi: 10.3389/fpubh.2021.695139] [Medline: 34395368]

19. MariaDB foundation homepage. MariaDB Foundation. URL: https://mariadb.org/ [accessed 2023-11-15]

20. Neo4j graph database and analytics. Neo4j. URL: https://neo4j.com/ [accessed 2023-11-15]

21. Neumann M, King D, Beltagy I, Ammar W. ScispaCy: fast and robust models for biomedical natural language processing. In: Proceedings of the 18th BioNLP Workshop and Shared Task. 2019. Presented at: BioNLP 2019; August 1, 2019; Florence, Italy. [doi: 10.18653/v1/w19-5034]

22. neo4j / graph-data-science. GitHub. URL: https://github.com/neo4j/graph-data-science [accessed 2023-11-15]

23. Pairo-Castineira E, Clohisey S, Klaric L, Bretherick AD, Rawlik K, Pasko D, GenOMICC Investigators, ISARIC4C Investigators, COVID-19 Human Genetics Initiative, 23andMe Investigators, BRACOVID Investigators, Gen-COVID Investigators, et al. Genetic mechanisms of critical illness in COVID-19. Nature. Mar 2021;591(7848):92-98. [doi: 10.1038/s41586-020-03065-y] [Medline: 33307546]

24. Teixeira PC, Dorneles GP, Santana Filho PC, da Silva IM, Schipper LL, Postiga IA, et al. Increased LPS levels coexist with systemic inflammation and result in monocyte activation in severe COVID-19 patients. Int Immunopharmacol. Nov 2021;100:108125. [FREE Full text] [doi: 10.1016/j.intimp.2021.108125] [Medline: 34543980]

25. Kratochvil MJ, Kaber G, Demirdjian S, Cai PC, Burgener EB, Nagy N, Stanford COVID-19 Biobank Study Group, et al. Biochemical, biophysical, and immunological characterization of respiratory secretions in severe SARS-CoV-2 infections. JCI Insight. Jun 22, 2022;7(12):e152629. [FREE Full text] [doi: 10.1172/jci.insight.152629] [Medline: 35730564]

26. Perico N, Cortinovis M, Suter F, Remuzzi G. Home as the new frontier for the treatment of COVID-19: the case for anti-inflammatory agents. Lancet Infect Dis. Jan 2023;23(1):e22-e33. [FREE Full text] [doi: 10.1016/S1473-3099(22)00433-9] [Medline: 36030796]

27. Consolaro E, Suter F, Rubis N, Pedroni S, Moroni C, Pastò E, et al. A home-treatment algorithm based on anti-inflammatory drugs to prevent hospitalization of patients with early COVID-19: a matched-cohort study (COVER 2). Front Med (Lausanne). Apr 22, 2022;9:785785. [FREE Full text] [doi: 10.3389/fmed.2022.785785] [Medline: 35530041]

28. Popovych V, Koshel I, Malofiichuk A, Pyletska L, Semeniuk A, Filippova O, et al. A randomized, open-label, multicenter, comparative study of therapeutic efficacy, safety and tolerability of BNO 1030 extract, containing marshmallow root, chamomile flowers, horsetail herb, walnut leaves, yarrow herb, oak bark, dandelion herb in the treatment of acute non-bacterial tonsillitis in children aged 6 to 18 years. Am J Otolaryngol. 2019;40(2):265-273. [FREE Full text] [doi: 10.1016/j.amjoto.2018.10.012] [Medline: 30554882]

29. Sava M, Sommer G, Daikeler T, Woischnig AK, Martinez AE, Leuzinger K, et al. Ninety-day outcome of patients with severe COVID-19 treated with tocilizumab - a single centre cohort study. Swiss Med Wkly. Aug 10, 2021;151:w20550. [FREE Full text] [doi: 10.4414/smw.2021.20550] [Medline: 34375986]

30. Patel SK, Juno JA, Lee WS, Wragg KM, Hogarth PM, Kent SJ, et al. Plasma ACE2 activity is persistently elevated following SARS-CoV-2 infection: implications for COVID-19 pathogenesis and consequences. Eur Respir J. May 13, 2021;57(5):2003730. [FREE Full text] [doi: 10.1183/13993003.03730-2020] [Medline: 33479113]

31. Yamaguchi T, Hoshizaki M, Minato T, Nirasawa S, Asaka M, Niiyama M, et al. ACE2-like carboxypeptidase B38-CAP protects from SARS-CoV-2-induced lung injury. Nat Commun. Nov 23, 2021;12(1):6791. [FREE Full text] [doi: 10.1038/s41467-021-27097-8] [Medline: 34815389]

32. Gupta S, Mitra A. Challenge of post-COVID era: management of cardiovascular complications in asymptomatic carriers of SARS-CoV-2. Heart Fail Rev. Jan 2022;27(1):239-249. [FREE Full text] [doi: 10.1007/s10741-021-10076-y] [Medline: 33426593]

33. Gargano JW, Wallace M, Hadler SC, Langley G, Su JR, Oster ME, et al. Use of mRNA COVID-19 vaccine after reports of myocarditis among vaccine recipients: update from the advisory committee on immunization practices - United States, June 2021. MMWR Morb Mortal Wkly Rep. Jul 09, 2021;70(27):977-982. [FREE Full text] [doi: 10.15585/mmwr.mm7027e2] [Medline: 34237049]

34. Oster ME, Shay DK, Su JR, Gee J, Creech CB, Broder KR, et al. Myocarditis cases reported after mRNA-based COVID-19 vaccination in the US from December 2020 to August 2021. JAMA. Jan 25, 2022;327(4):331-340. [FREE Full text] [doi: 10.1001/jama.2021.24110] [Medline: 35076665]

35. Lee E, Kim K, Kim M, Yang H, Yum HY, Lee M, et al. Adverse reactions to coronavirus disease 2019 vaccines in children and adolescents. Allergy Asthma Respir Dis. Jan 2022;10(1):9-14. [doi: 10.4168/aard.2022.10.1.9]

36. Zhu Y, Qin L, Yu JX, Cheng H. Finding top-k similar graphs in graph databases. In: Proceedings of the 15th International Conference on Extending Database Technology. 2012. Presented at: EDBT '12; March 27-30, 2012; Berlin, Germany. [doi: 10.1145/2247596.2247650]

37. Nikolentzos G, Meladianos P, Rousseau F, Stavrakas Y, Vazirgiannis M. Shortest-path graph kernels for document similarity. In: Proceedings of the 2017 Conference on Empirical Methods in Natural Language Processing. 2017. Presented at: EMNLP 2017; September 9-11, 2017; Copenhagen, Denmark. [doi: 10.18653/v1/d17-1202]

38. Han J, Sarica S, Shi F, Luo J. Semantic networks for engineering design: state of the art and future directions. J Mech Des. Feb 2022;144(2):020802. [doi: 10.1115/1.4052148]

39. Shi F, Chen L, Han J, Childs P. A data-driven text mining and semantic network analysis for design information retrieval. J Mech Des. Nov 2017;139(11):111402. [doi: 10.1115/1.4037649]

40. Wang JZ, Zhang Y, Dong L, Li L, Srimani PK, Yu PS. G-Bean: an ontology-graph based web tool for biomedical literature retrieval. BMC Bioinform. Nov 6, 2014;15(Suppl 12):S1. [doi: 10.1186/1471-2105-15-s12-s1]

41. Taduri S, Law KH, Kesan JP, Sriram RD. Utilization of bio-ontologies for enhancing patent information retrieval. In: Proceedings of the IEEE 43rd Annual Computer Software and Applications Conference. 2019. Presented at: COMPSAC 2019; July 15-19, 2019; Milwaukee, WI. [doi: 10.1109/compsac.2019.10189]

42. Dinh D, Tamine L. Identification of concept domains and its application in biomedical information retrieval. Inf Syst E Bus Manage. Sep 20, 2014;13:647-672. [doi: 10.1007/s10257-014-0259-y]

43. Gao S, Kotevska O, Sorokine A, Christian JB. A pre-training and self-training approach for biomedical named entity recognition. PLoS One. Feb 9, 2021;16(2):e0246310. [FREE Full text] [doi: 10.1371/journal.pone.0246310] [Medline: 33561139]

44. Wang X, Huang Z, van Harmelen F. Ontology-based semantic similarity approach for biomedical dataset retrieval. In: Proceedings of the Health Information Science. 2020. Presented at: HIS 2020; October 20-23, 2020; Amsterdam, The Netherlands. [doi: 10.1177/0002764289032005003]

45. Maraver P, Armañanzas R, Gillette TA, Ascoli GA. PaperBot: open-source web-based search and metadata organization of scientific literature. BMC Bioinformatics. Jan 24, 2019;20(1):50. [FREE Full text] [doi: 10.1186/s12859-019-2613-z] [Medline: 30678631]

46. Díaz-Galiano MC, Martín-Valdivia MT, Ureña-López LA. Query expansion with a medical ontology to improve a multimodal information retrieval system. Comput Biol Med. Apr 2009;39(4):396-403. [doi: 10.1016/j.compbiomed.2009.01.012] [Medline: 19268924]

47. Dong L, Srimani PK, Wang JZ. Ontology graph based query expansion for biomedical information retrieval. In: Proceedings of the IEEE International Conference on Bioinformatics and Biomedicine. 2011. Presented at: BIBM 2011; November 12-15, 2011; Atlanta, GA. [doi: 10.1109/bibm.2011.15]

48. Mohan S, Li D. MedMentions: a large biomedical corpus annotated with UMLS concepts. arXiv. Preprint posted online on February 25, 2019. [FREE Full text] [doi: 10.24432/C5G59C]

49. Basaldella M, Liu F, Shareghi E, Collier N. COMETA: a corpus for medical entity linking in the social media. In: Proceedings of the 2020 Conference on Empirical Methods in Natural Language Processing. 2020. Presented at: EMNLP 2020; November 16-20, 2020; Online. [doi: 10.18653/v1/2020.emnlp-main.253]

50. Fei H, Ren Y, Zhang Y, Ji D, Liang X. Enriching contextualized language model from knowledge graph for biomedical information extraction. Brief Bioinform. May 20, 2021;22(3):bbaa110. [doi: 10.1093/bib/bbaa110] [Medline: 32591802]

51. Kalyan KS, Rajasekharan A, Sangeetha S. AMMU: a survey of transformer-based biomedical pretrained language models. J Biomed Inform. Feb 2022;126:103982. [FREE Full text] [doi: 10.1016/j.jbi.2021.103982] [Medline: 34974190]

52. Welcome to the COVID-19 portfolio. National Institute for Health Office of Portfolio Analysis. URL: https://icite.od.nih.gov/covid19/search/ [accessed 2023-11-15]

53. Human coronaviruses data initiative. Lens. URL: https://about.lens.org/covid-19/ [accessed 2023-11-15]

54. Dagdelen J, Trewartha A, Huo H, Fei Y, He T, Cruse K, et al. COVIDScholar: an automated COVID-19 research aggregation and analysis platform. PLoS One. Feb 1, 2023;18(2):e0281147. [FREE Full text] [doi: 10.1371/journal.pone.0281147] [Medline: 36724184]

XSL·FO

RenderX

55. Wang X, Song X, Li B, Guan Y, Han J. Comprehensive named entity recognition on CORD-19 with distant or weak supervision. arXiv. Preprint posted online on March 27, 2020. [FREE Full text] [doi: 10.48550/arXiv.2003.12218]

56. Chen C, Ross KE, Gavali S, Cowart JE, Wu CH. COVID-19 Knowledge Graph from semantic integration of biomedical literature and databases. Bioinformatics. Dec 07, 2021;37(23):4597-4598. [FREE Full text] [doi: 10.1093/bioinformatics/btab694] [Medline: 34613368]

57. Steenwinckel B, Vandewiele G, Rausch I, Heyvaert P, Taelman R, Colpaert P, et al. Facilitating the analysis of COVID-19 literature through a knowledge graph. In: Proceedings of the 19th International Semantic Web Conference. 2020. Presented at: ISWC 2020; November 2-6, 2020; Athens, Greece. [doi: 10.1007/978-3-030-62466-8_22]

58. Pestryakova S, Vollmers D, Sherif MA, Heindorf S, Saleem M, Moussallem D, et al. COVIDPUBGRAPH: a FAIR knowledge graph of COVID-19 publications. Sci Data. Jul 08, 2022;9(1):389. [FREE Full text] [doi: 10.1038/s41597-022-01298-2] [Medline: 35803947]

59. Gütebier L, Bleimehl T, Henkel R, Munro J, Müller S, Morgner A, et al. CovidGraph: a graph to fight COVID-19. Bioinformatics. Oct 14, 2022;38(20):4843-4845. [FREE Full text] [doi: 10.1093/bioinformatics/btac592] [Medline: 36040169]

60. Roberts K, Alam T, Bedrick S, Demner-Fushman D, Lo K, Soboroff I, et al. Searching for scientific evidence in a pandemic: an overview of TREC-COVID. J Biomed Inform. Sep 2021;121:103865. [FREE Full text] [doi: 10.1016/j.jbi.2021.103865] [Medline: 34245913]

61. Chen Q, Allot A, Leaman R, Wei CH, Aghaarabi E, Guerrerio JJ, et al. LitCovid in 2022: an information resource for the COVID-19 literature. Nucleic Acids Res. Jan 06, 2023;51(D1):D1512-D1518. [FREE Full text] [doi: 10.1093/nar/gkac1005] [Medline: 36350613]

62. Full-text index overview. MariaDB. URL: https://mariadb.com/kb/en/full-text-index-overview/ [accessed 2023-11-15]

63. Fudolig MI, Alshaabi T, Arnold MV, Danforth CM, Dodds PS. Sentiment and structure in word co-occurrence networks on Twitter. Appl Netw Sci. Feb 14, 2022;7(1):1-27. [doi: 10.1007/s1109-022-00446-2]

64. Segev E. Textual network analysis: detecting prevailing themes and biases in international news and social media. Sociol Compass. Feb 14, 2020;14(4):e12779. [doi: 10.1111/soc4.12779]

65. Millington T, Luz S. Analysis and classification of word co-occurrence networks from Alzheimer's patients and controls. Front Comput Sci. Apr 29, 2021;3:1-12. [doi: 10.3389/fcomp.2021.649508]

66. Singhal A, Simmons M, Lu Z. Text mining genotype-phenotype relationships from biomedical literature for database curation and precision medicine. PLoS Comput Biol. Nov 30, 2016;12(11):e1005017. [FREE Full text] [doi: 10.1371/journal.pcbi.1005017] [Medline: 27902695]

67. Serna García G, Al Khalaf R, Invernici F, Ceri S, Bernasconi A. CoVEffect: interactive system for mining the effects of SARS-CoV-2 mutations and variants based on deep learning. Gigascience. Dec 28, 2022;12:giad036. [FREE Full text] [doi: 10.1093/gigascience/giad036] [Medline: 37222749]

68. frinve/graph-search. Docker. URL: https://hub.docker.com/r/frinve/graph-search [accessed 2024-05-14]

69. GRAPH-SEARCH homepage. GRAPH-SEARCH. URL: http://geco.deib.polimi.it/graph-search/ [accessed 2024-05-14]

70. FrInve / graph-search. GitHub. URL: https://github.com/FrInve/graph-search/wiki/ [accessed 2024-05-14]

## Abbreviations

**ACE2:** angiotensin-converting enzyme 2
**API:** application programming interface
**CIDO:** Coronavirus Infectious Disease Ontology
**CORD-19:** COVID-19 Open Research Dataset
**NER:** named entity recognition
**NPMI:** normalized pointwise mutual information
**NSAID:** nonsteroidal anti-inflammatory drugs
**UC:** use case
**UMLS:** Unified Medical Language System

XSL·FO
RenderX